\documentclass[prb,showpacs,aps,twocolumn]{revtex4} 
\usepackage{amsmath}
\usepackage{graphicx}
\usepackage{dcolumn}

\begin{document}

\title{Vortex configurations and critical parameters in superconducting thin films
containing antidot arrays: Nonlinear Ginzburg-Landau theory}
\author{G. R. Berdiyorov}
\email{golibjon.berdiyorov@ua.ac.be}
\author{M. V. Milo\v{s}evi\'{c}}
\author{F. M. Peeters}
\email{francois.peeters@ua.ac.be}

\affiliation{Departement Fysica, Universiteit Antwerpen,\\
Groenenborgerlaan 171, B-2020 Antwerpen, Belgium}

\date{\today}

\begin{abstract}

Using the non-linear Ginzburg-Landau (GL) theory, we obtain the
possible vortex configurations in superconducting thin films
containing a square lattice of antidots. The equilibrium
structural phase diagram is constructed which gives the different
ground-state vortex configurations as function of the size and
periodicity of the antidots for a given effective GL parameter
$\kappa^{*}$. Giant-vortex states, combination of giant- and
multi-vortex states, as well as symmetry imposed vortex-antivortex
states are found to be the ground state for particular geometrical
parameters of the sample. The antidot occupation number $n_o$ is
calculated as a function of related parameters and comparison with
existing expressions for the saturation number $n_s$ and with
experimental results is given. For a small radius of antidots a
triangular vortex lattice is obtained, where some of the vortices
are pinned by the antidots and some of them are located between
them. Transition between the square pinned and triangular vortex
lattices is given for different values of the applied field. The
enhanced critical current at integer and rational matching fields
is found, where the level of enhancement at given magnetic field
directly depends on the vortex-occupation number of the antidots.
For certain parameters of the antidot lattice and/or temperature
the critical current is found to be larger for higher magnetic
fields. Superconducting/normal $H-T$ phase boundary exhibits
different regimes as antidots are made larger, and we transit from
a plain superconducting film to a thin-wire superconducting
network. Presented results are in good agreement with available
experiments and suggest possible new experiments.

\end{abstract}


\maketitle

\section{Introduction}

Superconducting (SC) samples with periodic arrays of pinning sites
have received much attention over the last decade. It is now well
established that these artificial pinning centers (i) hold great
potential for enhancing the critical parameters of the sample and
(ii) give rise to different kinds of vortex behavior that is not
observed in the presence of random pinning. In this respect,
arrays of microholes (antidots)
\cite{harada,baert,MoshPRB96,MoshPRB98,metl,vanlook,silha2,silha,silha05,read,grigor,doria}
and submicron magnetic dots, \cite{Martin1,Martin2,mishko} have
been studied, as their presence in the SC film strongly modifies
the vortex structure compared to the one in non-patterned
films.~\cite{abrikos,abri_2}

Direct imaging experiments, \cite{harada} magnetization and
transport measurements, \cite{baert,MoshPRB96,MoshPRB98,metl} and
theoretical simulations
\cite{ReiPRB97,ReiPRB98,ReiPRB01,ReiPRB011,zhu} of vortex
structures in samples with periodic pinning centers have shown
that the vortices form highly ordered configurations at integer
$H_{n}=n\Phi_{0}/S$ and at some fractional
$H_{p/q}=\frac{p}{q}\Phi_{0}/S$ (n,p,q being integers) matching
fields, where $\Phi_{0}=hc/2e=2.07\cdot10^{-7} Gcm^{2}$ is the
flux quantum, and $S$ is the area of the primitive cell of the
artificial lattice. This remarkable variety of stabilized vortex
lattices may even be broadened by multiple possible degeneracies.
These commensurability effects between the pinning array and the
vortex lattice are responsible for an enhanced pinning and
consequently increased critical currents. Very recently Karapetrov
\textit{et al}.~\cite{STM} investigated vortex configurations in a
single crystal superconducting heterostructures with an array of
submicron normal metal islands by scanning tunneling microscopy.
They observed the coexistence of strongly interacting multiquanta
vortex lattice with interstitial Abrikosov vortices. Different
vortex phase transitions are given, which occur when the number of
magnetic flux quanta in the sample changes.


Motivated by those experimental studies on perforated
superconductors, significant efforts have been made on the
theoretical side as well. For example, extensive molecular
dynamics simulations
\cite{ReiPRB97,ReiPRB98,ReiPRB01,ReiPRB011,zhu} in the London
limit have been performed in an attempt to calculate the vortex
structure and their dynamics in a periodic pinning potential.
Although the general behavior of vortex lattices was accurately
described, made approximations are valid only in certain range of
parameters. Namely, in the London approach, vortices are
considered as classical point-particles (with different models for
their interaction) and the influence of the antidots is introduced
through model hole potential, which in principle can never be
generalized. Recently, Nordborg and Vinokur~\cite{nord} discussed
in the detail interaction of vortices with an arbitrarily large
cavity, but still within the London theory. This study was
actually an extension of the work of Mkrtchyan and
Shmidt,~\cite{Mkrt} who crudely estimated the maximum possible
number of vortices trapped by a single insulating inclusion with
radius $R$ as $n_{s}\cong R/2\xi(T)$, where $\xi(T)$ is the
temperature dependent coherent length. For regular arrays of
pinning centers the saturation number becomes $n_{s}\sim
(R/\xi(0))^{2}$ due to the vortex-vortex
interactions.~\cite{DoriaPC} The antidot-vortex interaction and
the following maximal occupation number of each antidot appear to
be crucial for many phenomena. For example, experiments on thin
films with a lattice of holes showed a ``localization transition''
\cite{bezr}: all vortices drop inside the holes when the coherence
length becomes larger than the interhole spacing. In Ref.
\cite{MoshPRB98} it was shown that the antidot size realizing the
optimum pinning is actually field-dependent. The effective
vortex-pinning potential and saturation number of the pinning
sites for different temperature and applied $dc$ fields were
recently investigated experimentally by means of
$ac$-susceptibility measurements, for superconducting films with
an array of antidots \cite{silha2,silha} and for the case of not
fully perforated holes (\textit{i.e.} blind holes).~\cite{read}

Most of the experiments on perforated superconducting films are
carried out in the effective type-II limit
($\kappa^{*}=\lambda^{2}/d\xi\gg 1/\sqrt{2}$, $d$ being the
thickness of the superconducting film and $\lambda$ the magnetic
penetration depth). In this regime, the vortices act like charged
point particles and their interaction with periodic pinning
potential can be described using molecular dynamic
simulations.~\cite{ReiPRB97,ReiPRB98,ReiPRB01,ReiPRB011}. However,
the overlap of vortex cores (with size $\sim \xi$), and the exact
shape of the inter-vortex interaction (depending on the
superconducting material properties reflected through $\kappa$),
may significantly modify the vortex structures and consequently
the critical current when this criteria is no longer satisfied.

Besides, the vortex-pinning and the critical current enhancement,
higher critical field ($H_{c3}$) near an open circular hole in a
thin film (the so-called ``surface superconductivity'') has been
predicted theoretically \cite{BuzdinPRB93} and confirmed
experimentally.~\cite{bezrJLTP} Cusps in the $H-T$ boundary were
observed, which occurs when the number of vortices which nucleate
inside the hole increases by one, similarly to the known
Little-Parks effect. The ratio between the critical fields in
perforated samples was estimated in limiting cases:
$H_{c3}/H_{c2}=1$ when $R\rightarrow 0$ (or $R<<\xi$) and
$H_{c3}/H_{c2}=1.695$ when $R\rightarrow \infty$.

In this work superconducting films with square arrays of antidots
are treated within the phenomenological Ginzburg-Landau theory.
This approach considers vortices as extended objects and no
approximations have to be made on e.g. the vortex-vortex
interaction and/or the vortex-antidot interaction. In Sec. II, the
details of our numerical formalism are given. Sec. III deals with
vortex lattices in perforated films in homogeneous magnetic field,
with emphasis on the number of pinned and interstitial vortices as
function of the antidot-size and interhole distance. In case of
weak pinning potentials, \textit{i.e.} small size anti-dots, we
discuss the triangle to square vortex lattice transition in Sec.
IV. In Sec. V, we address the behavior of critical current in the
sample as function of the applied field, for different geometrical
parameters, and temperature. The dependence of the critical field
on temperature, and different regimes in the $H-T$ phase diagram
are discussed in Sec. VI for different antidot-size. All presented
findings are then summarized in Sec. VII.


\section{Theoretical formalism}

\begin{figure}\centering
\vspace{0cm}
\includegraphics[scale=0.8]{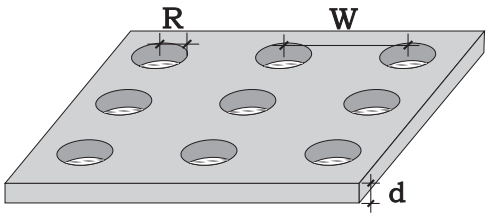}
\vspace{0cm} \caption{Schematic view of the sample: a
superconducting film (thickness $d$) with a regular array (period
$W$) of circular antidots (radius $R$).} \label{fig.1}
\end{figure}

In this work, we consider a thin superconducting film (of
thickness $d\ll \xi, \lambda$) with a square array of holes
(radius $R$, period $W$) immersed in an insulating media in the
presence of a perpendicular uniform applied field $H$ (see
Fig.~\ref{fig.1}). To describe the superconducting state of the
sample we solve the coupled nonlinear GL equations, which are
written in dimensionless units in the following
form~\cite{SchPRL98,SchPRB98}:

\begin{equation}
\left( -i\vec{\nabla}-\vec{A}\right) ^{2}\Psi =\Psi \left( 1-|\Psi
|^{2}\right),
 \label{GL1}
\end{equation}

\begin{equation}
-\kappa^{*}\Delta \vec{A}=\frac{1}{2i}\left( \Psi ^{\ast
}\vec{\nabla}\Psi -\Psi \vec{\nabla}\Psi ^{\ast }\right) -\left|
\Psi \right| ^{2}\vec{A}. \label{GL2}
\end{equation}
We measure the distance in units of the coherence length $\xi$,
the vector potential $\vec{A}$ in $c\hbar/2e\xi$, the magnetic
field $H$ in $H_{c2}=c\hbar/2e\xi^{2}=\kappa\sqrt{2}H_{c}$, and
the order parameter $\Psi$ in $\sqrt{-\alpha/\beta}$ with
$\alpha$, $\beta$ being the GL coefficients.

The magnitude of the applied magnetic field $H=n\Phi_{0}/S$ is
determined by the number $n$ of flux quanta
$\Phi_{0}=hc/2e=2.07\cdot 10^{-7}$Gcm$^{2}$ piercing through the
rectangular simulation area $S=W_{x}\times W_{y}=N_xN_yW^2$, with
$W_{x(y)}=N_{x(y)}W$ and $N_x$ and $N_y$ are integers. At the
superconductor/insulator interface we impose the boundary
condition corresponding to zero normal component of the
superconducting current. The periodic boundary conditions for
$\vec{A}$ and $\Psi$ (simulating the periodicity of both
superconducting film and antidot lattice) have the
form~\cite{DoriaBC}
\begin{equation}
\vec{A}(\vec{\rho}+\vec{b}_{i})=\vec{A}(\vec{\rho})+
\vec{\nabla}\eta_{i}(\vec{\rho}),
 \label{b1}
\end{equation}
\begin{equation}
\Psi(\vec{\rho}+\vec{b}_{i})=\Psi\cdot exp(2\pi i
\eta_{i}(\vec{\rho})/\Phi_{0}),
 \label{b2}
\end{equation}
where $\vec{b}_{i}$ ($i=x,y$) are lattice vectors, and $\eta_{i}$
is the gauge potential. These boundary conditions imply that
$\vec{A}$, $\Psi$ are invariant under lattice translations
combined with specific gauge transformations $\eta_{x,y}$. Other
quantities, such as the magnetic field, the current or the order
parameter density are periodic. We use the Landau gauge
$\vec{A}_{0}=Hx\vec{e}_{y}$ for the external vector potential and
$\eta_{x}=HW_{x}y+C_{x}$, $\eta_{y}=C_{y}$, with $C_{x}$, $C_{y}$
being constants.~\cite{mishko} Without antidot lattice, when the
film is invariant under infinitely small translations, the free
energy does not depend on $C_{x}$, $C_{y}$. The vortex lattice is
only shifted relative to the simulation region when $C_{x}$ and
$C_{y}$ are varied. This is not the case for an antidot lattice,
when the change of these parameters leads to a displacement of the
vortex lattice relative to the holes, leading to a variation of
the free energy. In general, one has to minimize the free energy
with respect to $C_{x}$, $C_{y}$. It can be shown that such a
minimization gives a zero current when averaged over the cell
area. We find that for a supercell having one hole, the optimal
values are given by $C_{x,y}=0,\pm \pi$ and that for the supercell
with $2^{N}$ holes the choice $C_{x,y}=0$ provides the minimum
free energy.

We solved the system of Eqs.~(\ref{GL1},\ref{GL2})
self-consistently using the link variable approach~\cite{Kato} in
a finite-difference representation of the order parameter and the
vector potential using a uniform cartesian space grid $(x,y)$. The
first GL equation is solved with a Gauss-Seidel iteration
procedure.~\cite{SchPRL98} The vector potential is then obtained
with the fast Fourier transform technique. The temperature is
indirectly included in the calculation through the temperature
dependence of the coherence length
$\xi(T)=\xi(0)/\sqrt{|1-T/T_{c0}|}$ and penetration depth
$\lambda(T)=\lambda(0)/\sqrt{|1-T/T_{c0}|}$, where $T_{c0}$ is the
critical temperature at zero magnetic field.


%

\section{Vortex lattices - influence of geometrical parameters}

\begin{figure}[t] \centering
\vspace{0cm}
\includegraphics[scale=0.55]{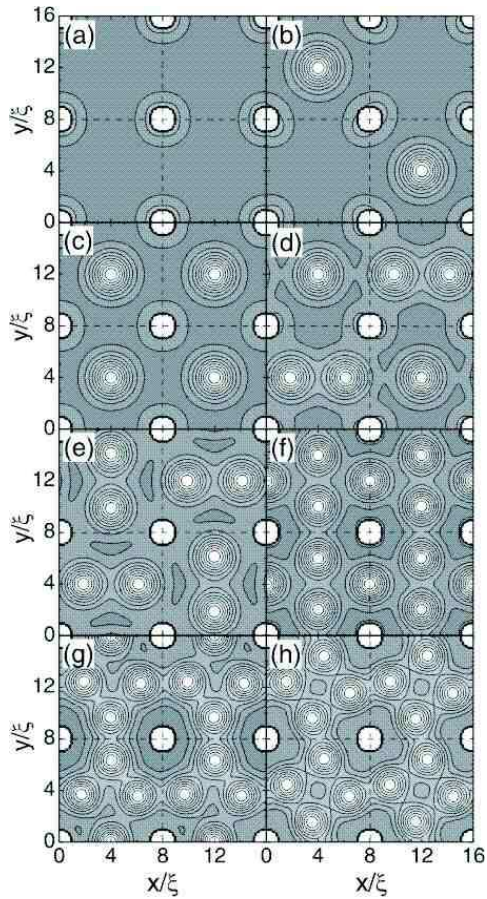}
\vspace{0cm} \caption{Contour plot of the Cooper-pair density
(white/dark color - low/high density) in the superconducting film
with antidots of radius $R=\xi$ and interhole distance $W=8.0\xi$
for the matching fields: $H_{n}=1$ (a), 3/2 (b), 2 (c), 5/2 (d), 3
(e), 4 (f), 9/2 (g), and 5 (h). The film thickness is $d=0.1\xi$
and the effective GL parameter $\kappa^{*}=10$. Dashed lines
indicate the antidot lattice.} \label{fig.2}
\end{figure}

We first consider a supercell containing four holes
($W_{x}=W_{y}=2W$) of radius $R$, with lattice period $W$ (see
Fig.~\ref{fig.1}). Although our approach is valid for any integer
number of flux-quanta piercing through the simulation region, we
will restrict ourselves here to the so-called (integer and
fractional) matching vorticities.


Figure \ref{fig.2} shows contour plots of $|\Psi|^2$ of the vortex
lattice in case of antidots of radius $R=\xi$ and interhole
distance $W=8.0\xi$ for different matching fields $H_{n}$. The
film thickness is $d=0.1\xi$ and the effective GL parameter
$\kappa^{*}=\lambda^{2}/d\xi=10$. At the first matching field all
vortices are trapped in the antidots [Fig. \ref{fig.2}(a)].
Because of their small radius, each antidot is able to pin only
one vortex, and additional vortices localize at interstitial sites
when $H>H_{1}$ [see Fig. \ref{fig.2}(b) for $H=H_{3/2}$]. At the
second matching field [Fig. \ref{fig.2}(c)] vortices occupy all
interstitial sites, forming again a square lattice. For
$H=H_{5/2}$, [Fig. \ref{fig.2}(d)] vortices form an ordered
lattice with an additional vortex at every other interstice, added
to the $H_{2}$ case. Note that the size of the vortex cores at
neighboring interstitial sites differs: two vortices at the same
interstice strongly interact and effectively bound each others
core-areas, while the neighboring single interstitial vortex does
not suffer from any lack of space resulting in its larger core. At
the third matching field (three vortices per unit cell) two
interstitial vortices in adjacent cells alternate in position [see
Fig. \ref{fig.2}(e)], preserving the two-fold symmetry, but the
vortex unit cell is 4 times the antidot lattice unit cell. At
$H=H_{4}$ [Fig. \ref{fig.2}(f)] we observe the first evidence of
the competition between the Abrikosov vortex lattice
(characteristic for thin film superconductors) and the symmetry of
the pinning lattice, as vortices (including pinned ones) form a
slightly deformed hexagonal lattice. Notice that for $H=H_{9/2}$
[Fig.~\ref{fig.2}(g)] the number of vortices per antidot unit cell
is $n=9/2$ and the vortex lattice unit cell is twice the antidot
unit cell. At the fifth matching field [Fig. \ref{fig.2}(h)] the
dense packing of vortices and their consequent strong interaction
with the antidot lattice result in the restoration of the square
lattice symmetry but the vortex lattice is tilted over
$35^{\circ}$ with respect to the antidot lattice. Our results are
in excellent agreement with the experiment of Ref. \cite{harada}
and previous molecular dynamics simulations,\cite{ReiPRB98} in
certain parameter range. Namely, the vortex configurations are
mainly determined by the pinning force of each antidot, and the
vortex-vortex interaction. The latter is very dependent on the
density of vortex packing, as the known expressions for the
vortex-vortex interaction do not take into account possible
overlap of vortex cores. On the other hand, the antidot-vortex
pinning potential is determined by the antidot-size and the period
of the antidot-lattice. Our approach takes all these aspects into
account and their influence will be discussed in the remainder of
this paper.

\begin{figure} \centering
\vspace{0cm}
\includegraphics[scale=0.55]{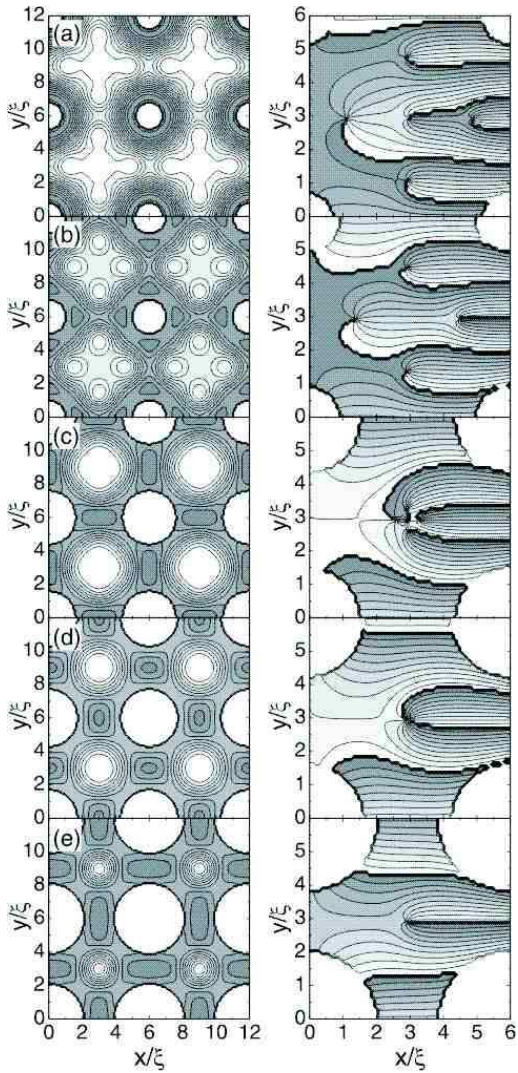}
\vspace{0.25cm} \caption{Contour plot of the Cooper-pair density
(left column) and phase of the order parameter (right column,
limited to a single antidot lattice unit) for the sixth matching
field $H=H_{6}$ and for different values of the hole radius:
$R=0.8\xi$ (a), $1\xi$ (b), $1.6\xi$ (c), $1.7\xi$ (d) and
$2.1\xi$ (e). The lattice period is $W=6\xi$ and the GL parameter
$\kappa^{*}=10$. } \label{fig.3}
\end{figure}
%

%
%

%
\begin{figure} \centering
\vspace{0cm}
\includegraphics[scale=0.65]{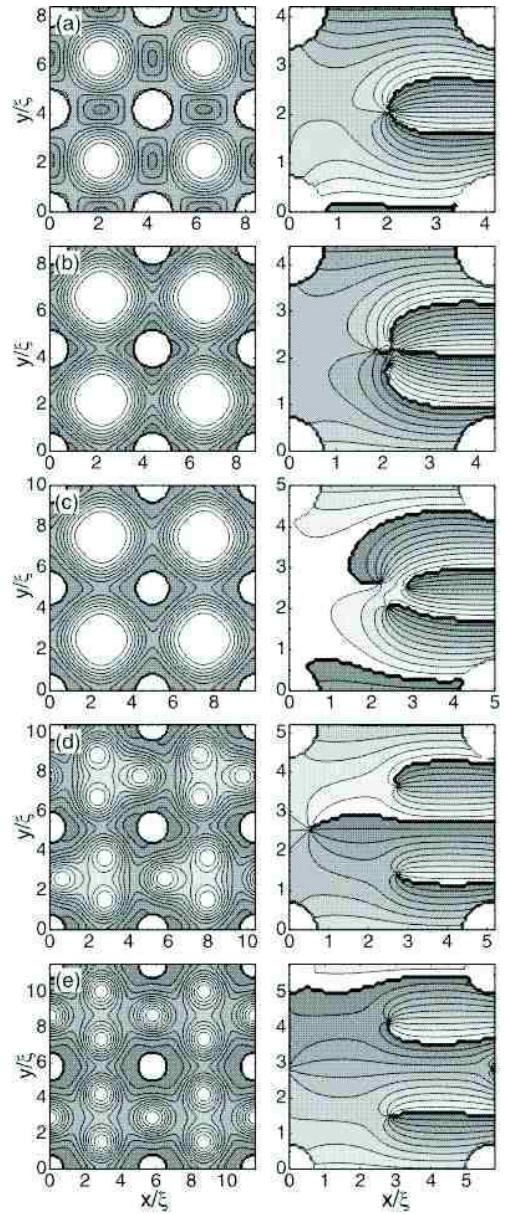}
\vspace{0cm} \caption{Contour plot of the Cooper-pair density
(left column) and phase of the order parameter (right column) for
the fourth matching field $H=H_{4}$ and for the different values
of the lattice period: $W=4.2\xi$ (a), $4.4\xi$ (b), $5\xi$ (c),
$5.2\xi$ (d) and $5.8\xi$ (e). The radius of the holes is
$R=0.8\xi$ and GL parameter $\kappa^{*}=10$. } \label{fig.4}
\end{figure}
\begin{figure}[t] \centering
\vspace{0cm}
\includegraphics[scale=0.5]{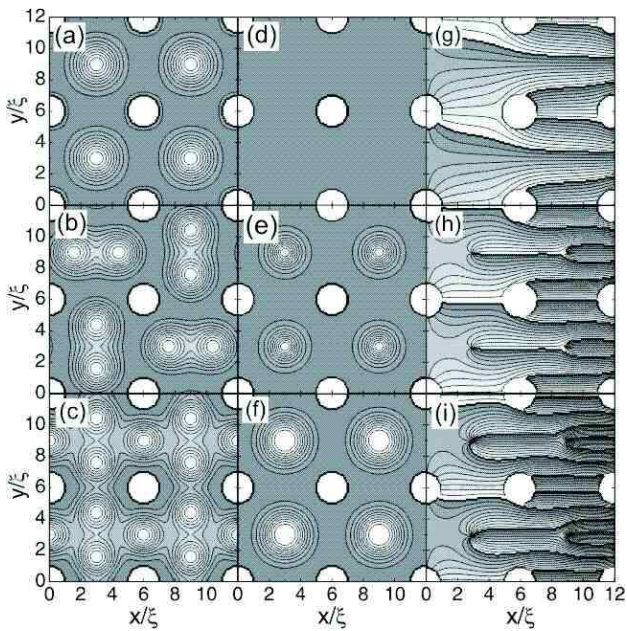}
\vspace{0.25cm} \caption{Contour plots of the Cooper-pair density
for $\kappa^{*}=10$ (a-c) and $\kappa^{*}=0.1$ (d-f). The lattice
period is $W=6\xi$ and the radius is $R=1\xi$. Figures (g-i) show
the phase of the order parameter of the states shown in (d-f). The
first row is for $H=H_2$, the second for $H=H_3$ and the bottom
row for $H=H_4$.} \label{fig.5}
\end{figure}
\begin{figure}[t] \centering
\vspace{0cm}
\includegraphics[scale=0.6]{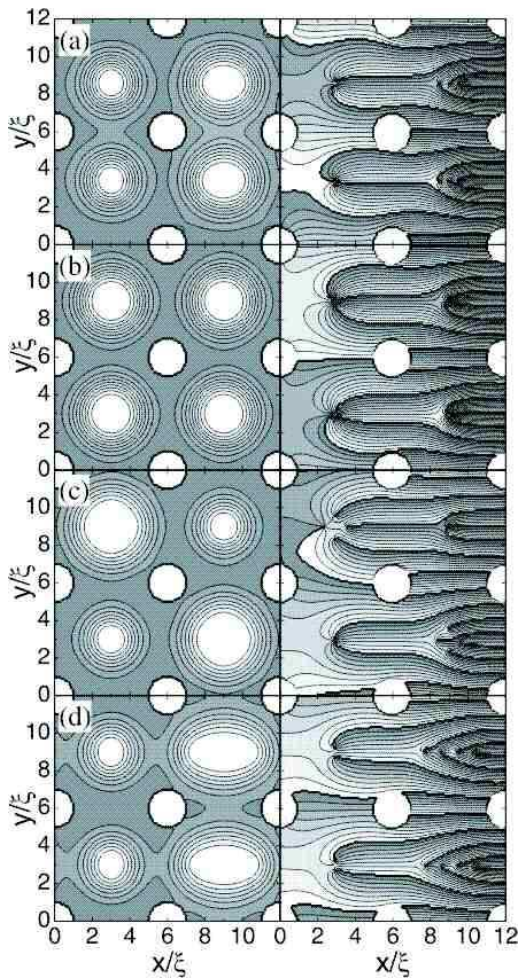}
\vspace{0.25cm} \caption{Contour plots of the Cooper-pair density
(left column) and the phase of the order parameter (right column)
of the sample in Fig.~\ref{fig.5} for different metastable vortex
states at $H=H_4$ and $\kappa^{*}=0.1$.} \label{fig.6}
\end{figure}
\begin{figure}[t] \centering
\vspace{0cm}
\includegraphics[scale=1.15]{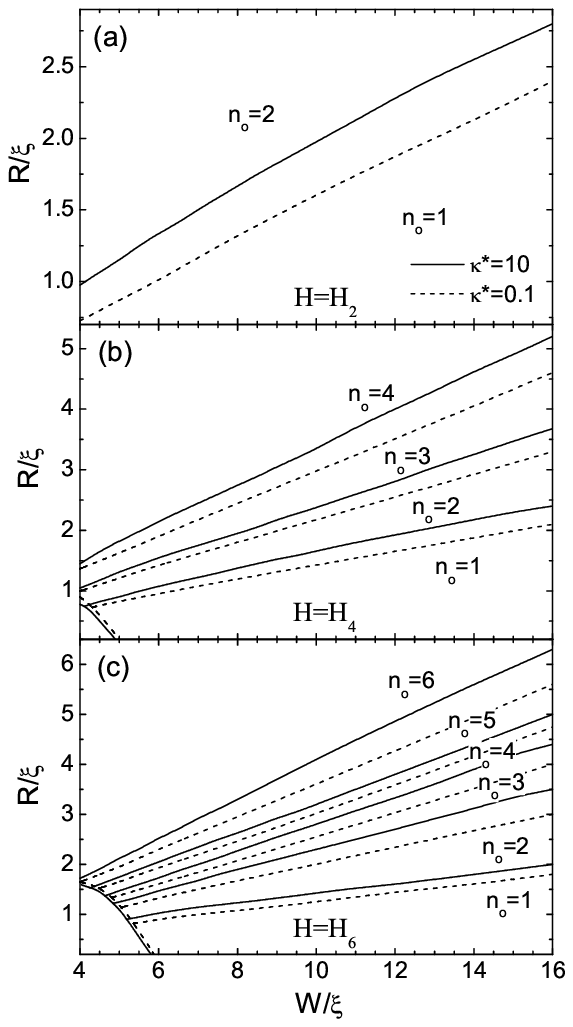}
\vspace{0cm} \caption{The vortex phase diagram: the dependence of
the hole occupation number $n_{o}$ on the radius of the holes $R$
and the distance between them $W$, for $\kappa^{*}=10$ (solid
curves) and $\kappa^{*}=0.1$ (dashed curves) at the matching
fields $H=H_{2}$ (a), $H=H_{4}$ (b) and $H=H_{6}$ (c). The gray
region shows the normal state region for $\kappa^{*}=10$
($\kappa^{*}=0.1$ when delimited by the full curve).}
\label{fig.7}
\end{figure}

As shown in Fig. \ref{fig.2}, for small hole radii, each antidot
pins only one vortex (the hole occupation number $n_o$, the number
of vortices sitting in the holes, equals 1), and the remaining
vortices reside between the holes. One expects that, for larger
hole radius $R$, vortex configurations with multi-quanta vortices
in each hole can become energetically
preferable.~\cite{STM,BuzdinPRB93} Figure~\ref{fig.3} shows the
contour plots of the Cooper-pair density at the sixth matching
field for different antidot radii. The antidot lattice period is
$W=6\xi$, the thickness is $d=0.1\xi$ and $\kappa^{*}=10$. The
number of vortices captured by each hole changes from one for
$R=0.8\xi$ [Fig. \ref{fig.3}(a)] to five for $R=2.1\xi$ [Fig.
\ref{fig.3}(e)]. The vortex arrangement outside the holes is
determined not only by their mutual interaction, but also by the
attraction with the antidots and the repulsion by their pinned
vortices. For small radius $R$ a multivortex structure is found at
the interstitial sites, as apparent from Fig.~\ref{fig.3}(a,b)
(see also Fig.~\ref{fig.2}). By further increasing $R$ some of the
vortices enter the holes and the remaining vortices are strongly
caged between the antidots, resulting in the formation of giant
vortices [Fig.~\ref{fig.3}(d)] and symmetry imposed
vortex-antivortex pairs~\cite{PRL06} [Fig.~\ref{fig.3}(c)]. This
is apparent from the contour plot of the phase of the order
parameter $\Psi$ (right column of Fig.~\ref{fig.3}).


Similar behavior can be achieved if the hole-size is kept the
same, but the period of the hole lattice is decreased, as
illustrated in Fig.~\ref{fig.4}. For small period $W$ the
interstitial vortices form a giant vortices [Fig.~\ref{fig.4}(a)]
because of the strong interaction with the pinned vortices in the
antidots. With increasing $W$ one extra vortex is depinned and,
due to the symmetry of the sample, vortex-antivortex pair with
four vortices and one antivortex is formed in each interstitial
site [Fig.~\ref{fig.4}(b)]. Further increase of $W$ leads to a
triangular vortex structure at the interstitial sides with chosen
orientation that minimizes the energy between neighboring cells
[Figs.~\ref{fig.4}(c-e)].

It is well known that the vortex-vortex interaction changes sing
at the point $\kappa=1/\sqrt{2}$. For $\kappa>1/\sqrt{2}$,
vortices repel each other while for $\kappa<1/\sqrt{2}$ they
attract. To see how this attractive interaction modifies the
different vortex lattice configurations we consider a sample with
small $\kappa$. Fig.~\ref{fig.5} shows the contour plots of the
Cooper-pair density for $\kappa^{*}=10$ (type-II regime) and
$\kappa^{*}=0.1$ (type-I regime) for the second, third and fourth
matching fields. For the given parameters of the sample and for
$\kappa^{*}=10$ each hole pins one vortex and the remaining
vortices sit at interstitial sites [Figs.~\ref{fig.5}(a-c)]. The
occupation number of each hole is increased to two in the type-I
sample [see Figs.~\ref{fig.5}(d-i)]) due to the enhanced expulsion
of the magnetic field by the superconductor. Moreover, because of
the attractive interaction between vortices, giant vortices become
energetically more favorable contrary to the case for
$\kappa^*=10$. Due to the instabilities of vortex states, which is
common for type-I superconductors, variety of metastable vortex
structures can be found (see also Ref.~\cite{PC06}). As an example
we show in Fig.~\ref{fig.6} different metastable vortex states of
the sample in Fig.~\ref{fig.5} for $\kappa^{*}=0.1$. The free
energies of those states are: $F/F_0=$-0.3268 (a), -0.2823 (b),
-0.2787 (c) and -0.2755 (d). The ground state free energy
[Fig.~\ref{fig.5}(f,i)] is $F/F_0=-0.3759$. Notice that because of
the attractive interaction a giant vortex state is always favored.


To summarise the above findings, we constructed the equilibrium
vortex phase diagram, which shows the dependence of
antidot-occupation number $n_{o}$ on $R$ and $W$ for two values of
$\kappa^{*}$, at the second, fourth and sixth matching fields
(Fig.~\ref{fig.7}). The ground- and metastable states are
determined in our calculation by comparing the energy of all
stable vortex states found when starting from different randomly
generated initial conditions. The procedure of finding the free
energy of the different metastable states was similar to that used
for the case of mesoscopic superconducting
disks.~\cite{SchPRL98,SchPRB98} It should be noted, that an
energetically unfavorable state remain stable in the wide range of
variation of $R$ and $W$. Therefore, the transitions between the
vortex states with different occupation numbers are of
first-order. It is seen from this figure that $n_o$ increases as
the applied field is increased, which is in agreement with
experimental results and theoretical
predictions.~\cite{baert,doria,STM,BuzdinPRB93} With decreasing
$\kappa^{*}$ the threshold hole-radius for capturing another
vortex decreases due to the smaller pinned vortex-interstitial
vortex repulsion.


%
\begin{figure} \centering
\vspace{0cm}
\includegraphics[scale=1.2]{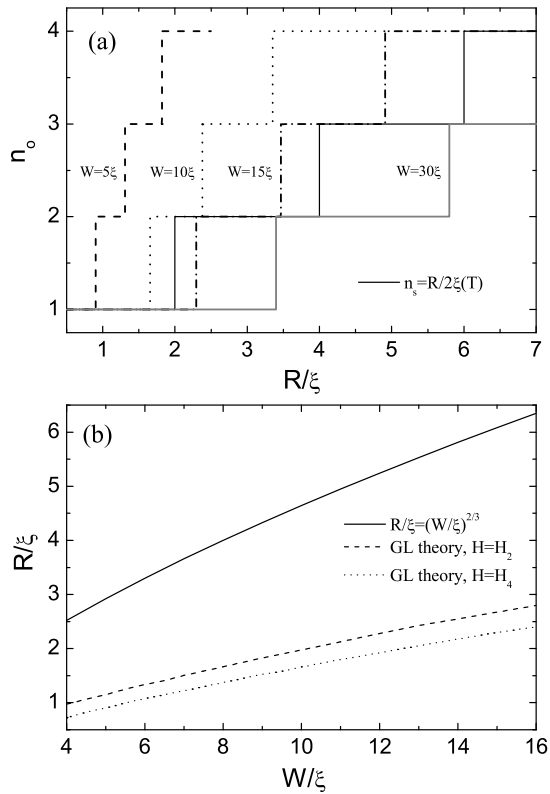}
\vspace{0cm} \caption {(a) The antidot occupation number as a
function of the antidot radius from the London theory~\cite{Mkrt}
$n_s=R/2\xi(T)$ (solid line) and from the GL theory for the period
$W=5\xi$ (dashed curve), $W=10\xi$ (dotted curve), $W=15\xi$
(dash-dotted curve) and $W=30\xi$ (gray curve) for the fourth
matching field and $\kappa^*=10$. (b) The critical hole radius
corresponding to $n_o=1\rightarrow n_o=2$ transition as a function
of the period $W$. Solid curve is obtained from the London
theory~\cite{BuzdinPRB93} and dashed (dotted) curve is the result
from the GL theory for $H=H_2$ ($H=H_4$) and $\kappa^*=10$.}
\label{fig.8}
\end{figure}

Let us compare our numerical results for the hole-occupation
number $n_{o}$ with existing theoretical predictions. The
saturation number $n_{s}$ ($n_{s}=n_{o}$ for larger fields) is
usually estimated as $n_{s}=R/2\xi(T)$.~\cite{Mkrt}
Fig.~\ref{fig.8}(a) shows the hole occupation number $n_o$
obtained from this expression and the one from our GL calculation
for different period of the antidots. It is seen from this figure
that this expression underestimates $n_o$ for small period $W$
(dashed and dotted curves). This is due to the fact that the last
expression does not account for the interaction between vortices
sitting at different holes. For larger period $W$ the occupation
number is smaller in our calculation for a given radius of the
holes (gray curve). A more accurate analysis was presented by
Buzdin \cite{BuzdinPRB93} for bulk superconductors within the
London approach. However, his estimation of the critical hole
radius $R/\xi\approx(W/\xi)^{2/3}$ (for $W\ll \lambda$) and
$R/\xi\approx\kappa^{2/3}$ (for $\lambda\ll W$) corresponding to
the transition from single flux-quantum to two flux-quanta
captured by the hole, differ from our numerically exact results,
\textit{i.e.}, the magnitude of the critical hole radius is
largely overestimated in Ref.~\cite{BuzdinPRB93} for both small
and large period $W$ [see Fig.~\ref{fig.8}(b)].

\begin{figure} \centering
\vspace{0cm}
\includegraphics[scale=1.15]{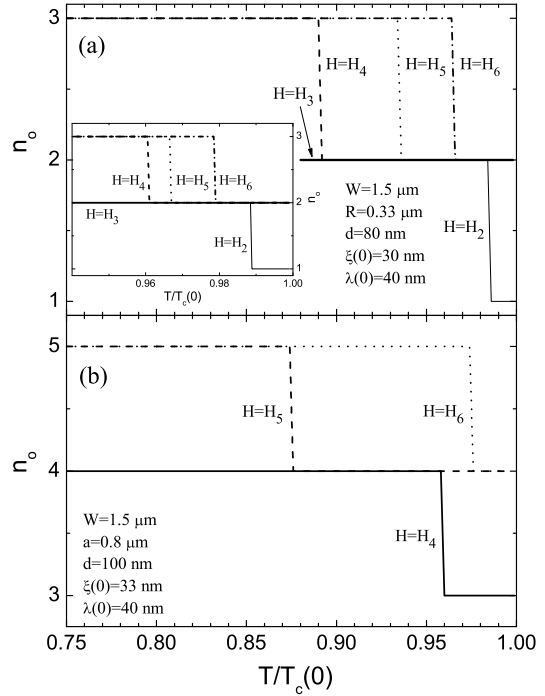}
\vspace{0cm} \caption{The antidot occupation number $n_{o}$ as a
function of temperature $T/T_{c0}$ for different matching fields
for circular antidots ($R=0.33\mu$m) (a) and square ($a=0.8\mu$m)
(b) holes. The lattice period for both samples is $W=1.5 \mu$m and
film thickness is $d=80$nm (a) and $d=100$nm (b). The inset in (a)
shows the hole occupation number $n_{o}$ of the same sample as a
function of temperature for a different temperature dependence of
$\xi(T)=\xi(0)\sqrt{|1-t^4|}/(1-t^2)$ and
$\kappa=\kappa(0)/(1+t^2)$, where $t=T/T_{c0}$ and
$\kappa(0)=\lambda(0)/\xi(0)$.} \label{fig.9}
\end{figure}

The maximum number of flux quanta that can be trapped in a pinning
center in a thin superconducting film was recently studied
experimentally using scanning Hall probe microscopy~\cite{grigor}
and ac susceptibility measurements.~\cite{silha} In the latter
case the saturation number was obtained from the transition to
different dynamic regimes, as the interstitial vortices have
higher mobility than those pinned by the antidots. They studied
thin Pb films containing a square antidot array of period
$d=1.5\mu$m. The antidots had circular (square) shape with radius
$R=330$nm (size $a=0.8\mu$m), the film thickness was $d=80$nm
($d=100$nm) and the coherence length at zero temperature was
estimated $\xi(0)=30$nm ($\xi(0)=33$nm) in Ref.~\cite{grigor}
(Ref.~\cite{silha}). Let us first discuss the results for the
sample of Ref.~\cite{grigor}, where the experimentally obtained
saturation number was $n_s$=2 at $T/T_{c0}=0.77$.
Fig.~\ref{fig.9}(a) shows the antidot occupation number $n_{o}$ as
a function of temperature for different applied matching fields.
At small applied fields ($H\leq H_3$) the occupation number is
equal to two, which is in agreement with the experimentally
obtained $n_s$. With increasing applied field $H>H_3$ one more
vortex is trapped by the holes, \textit{i.e.} $n_o=3$, which is
now larger than the experimental value. At higher temperatures
$T>0.89T_{c0}$, $n_o$ again becomes equal to two. In this case one
would estimate the saturation number from $n_s\approx
R/2\xi(T)$~\cite{Mkrt} to be $n_s=1$ for $0.967T_{c0}<T<T_{c0}$
and $n_s=2$ for $0.868T_{c0}<T<0.967T_{c0}$. We found the
occupation number equal to $n_o=1$ only for the second matching
field at the temperature range $0.985T_{c0}<T<T_c$. The estimation
of Buzdin~\cite{BuzdinPRB93} for the critical hole radius
$R^3<\xi(T)\lambda(T)^2$, where the transition from $n_o=1$ to
$n_o=2$ occurs, gives the temperature range $T<0.985T_{c0}$. We
found this transition at this temperature only for the second
matching field. For larger fields the occupation number is always
larger than unity. The giant vortex state is found only at $H=H_4$
for $T>0.984T_{c0}$ and the vortex-antivortex state is formed at
$H=H_5$ for the temperatures $T>0.986T_{c0}$.

Up to now we use the temperature dependence for the coherence
length and penetration depth as shown at the end of Sec. II which
is obtained from the BCS theory~\cite{tinkham} and is valid near
$T_c$. In this case the GL parameter $\kappa$ is temperature
independent. In recent experiments on Pb arrays of nanowires
arrays Stenuit \textit{et al.}~\cite{wire} found that the
following temperature dependence of the coherence length
$\xi(T)=\xi(0)\sqrt{|1-t^4|}/(1-t^2)$ and penetration depth
$\lambda(T)=\lambda(0)/\sqrt{1-t^4}$, which leads to a temperature
dependence of the GL parameter $\kappa=\kappa(0)/(1+t^2)$, agrees
better with experiment. Here $t=T/T_{c0}$ and
$\kappa(0)=\lambda(0)/\xi(0)$. These expressions are obtained from
the two-fluid model. We calculated the hole occupation number
$n_o$ using the above temperature dependencies, which is shown in
the inset of Fig.~\ref{fig.9}(a). It is seen from this figure that
the transition from $n_o=n$ to $n_o=n$-1 occurs now at higher
temperatures, but the results are qualitatively similar with the
earlier results.

Fig.~\ref{fig.9}(b) shows the occupation number $n_o$ as a
function of temperature for different matching fields for the
sample of Ref.~\cite{silha}. The experimentally obtained
saturation number was $n_s=3$ for temperatures $T/T_{c0}>0.97$.
Our calculations give the same occupation number $n_o=3$ for this
range of temperatures but only for fields $H\leq H_4$. The
expression for the saturation number $n_s\approx a/4\xi(T)$ gives
in this case $n_s=1$. At larger fields the occupation number
increases, but still there will be interstitial vortices in the
sample.~\cite{STM} These interstitial vortices lead to a larger
dissipation in the sample which was used as the criterium for the
determination of the saturation number $n_s$. But our calculations
show that the appearance of interstitial vortices does not
indicate the saturation of trapped vortices.


%
%


%
%

%
\begin{figure}[t] \centering
\vspace{0cm}
\includegraphics[scale=0.55]{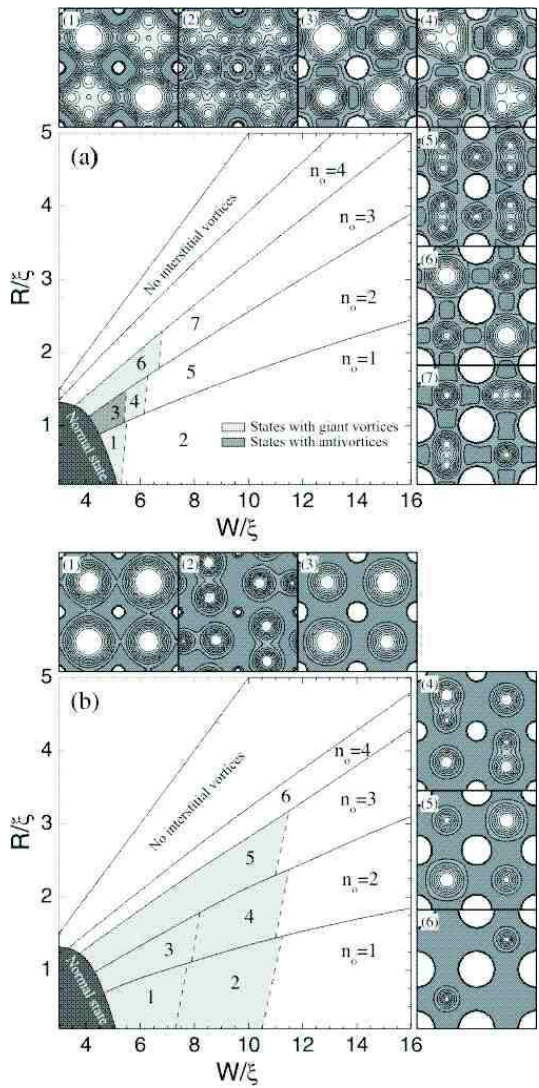}
\vspace{0cm} \caption{The ground-state vortex lattice at
$H=H_{9/2}$ as a function of the antidot-radius $R$ and their
periodicity $W$, for $\kappa^*=10$ (a) and $\kappa^*=0.1$ (b). The
solid lines denote the first order transitions between the states
with different antidot-occupation number $n_o$, and dashed ones
depict second order configurational transitions. The insets show
the Cooper-pair density plots of the corresponding states
indicated by the numbers in the phase diagram.} \label{fig.10}
\end{figure}

We have shown in our recent paper~\cite{PRL06} that a rich variety
of ordered vortex structures: a combination of giant vortices with
mltivortices and vortex-antivortex pairs are found in perforated
superconducting samples for fractional matching fields. Here we
consider the dependence of these vortex states on the effective GL
parameter $\kappa^*$. As an example we constructed the equilibrium
vortex phase diagram for $H=H_{9/2}$ rational matching field (i.e.
4.5 flux quanta per antidot) as a function of $R$ and $W$, and for
$\kappa^*=10$ [Fig. \ref{fig.10}(a)] and $\kappa^*=0.1$ [Fig.
\ref{fig.10}(b)]. In type-II regime [Fig. \ref{fig.10}(a)], for
larger period $W$ the vortex configuration always consists of
individual vortices, except for $n_o$ flux lines pinned by each
hole. One extra vortex per two holes is shared between the
adjacent cells (insets 2, 5) or situated in every other cell
(inset 7). With decreasing $W$, the interstitial vortices become
strongly caged between the neighboring antidots, resulting in the
distortion of the individual-vortex lattice. In this case, the
$n$-th matching field becomes larger than the second critical
field for $W/\xi<\sqrt{2\pi n}$. But the superconducting state in
perforated films still survives due to enhanced superconductivity
in close proximity around the holes (due to surface
superconductivity). If the radius of the holes is then increased,
individual vortices captured at interstitial sites can merge for
$H>H_{c2}$ to a giant vortex [(region 1 in Fig.~\ref{fig.10}(a)].
This transition does not show any hysteretic behavior and is,
therefore, of second order (similarly to the case for mesoscopic
disks \cite{SchPRL98}). The creation of these giant vortex states
is favored because of the repulsion of the vortices by the
supercurrents around the holes (and consequent compression of
vortices in the central part of the interstitial regions).

The influence of caging depends on the number of confined
vortices; the combinations of giant- and multi-vortices may be
formed in the interstitial sites (insets 1, 4 and 6). For
interstitial vorticity 3, a vortex-antivortex pair may nucleate,
so that local vortex structure conforms with the square symmetry
of the pinning lattice [see inset (3) in Fig.~\ref{fig.10}(a) and
Ref. \cite{PRL06}].

Fig. \ref{fig.10}(b) shows the ground-state phase diagram found
for $\kappa^*=0.1$. Compared to Fig. \ref{fig.10}(a), the
threshold antidot-radius for capturing another vortex decreases
due to the enhanced screening of the applied field. Due to the
attractive vortex-vortex interaction in type-I samples,
giant-vortex states become energetically favorable at the
interstitial sites and spread over the majority of the $W-R$ phase
diagram (light gray areas). For a dense antidot lattice, giant
vortices with different vorticity are found in adjacent cells
[$L=3$ and $L=4$ (inset 1), and $L=2$ and $L=3$ (inset 3)].
Contrary to the type-II case, these giant vortices can split to
smaller giant vortices for larger spacing of antidots. They
exhibit single-vortex behavior, forming the lattice of 2-quanta
and single-quanta vortices (insets 2 and 4). Such new
quasi-Abrikosov lattices of giant-vortices result from the
competition of vortex-vortex attraction and imposed square
symmetry of pinning. At the same time, these competing
interactions cause the complete disappearance of the
vortex-antivortex structures as found in type-II samples.

\section{Triangular to square vortex lattice transition in the presence of a square antidot lattice}


%
\begin{figure} \centering
\vspace{0cm}
\includegraphics[scale=0.5]{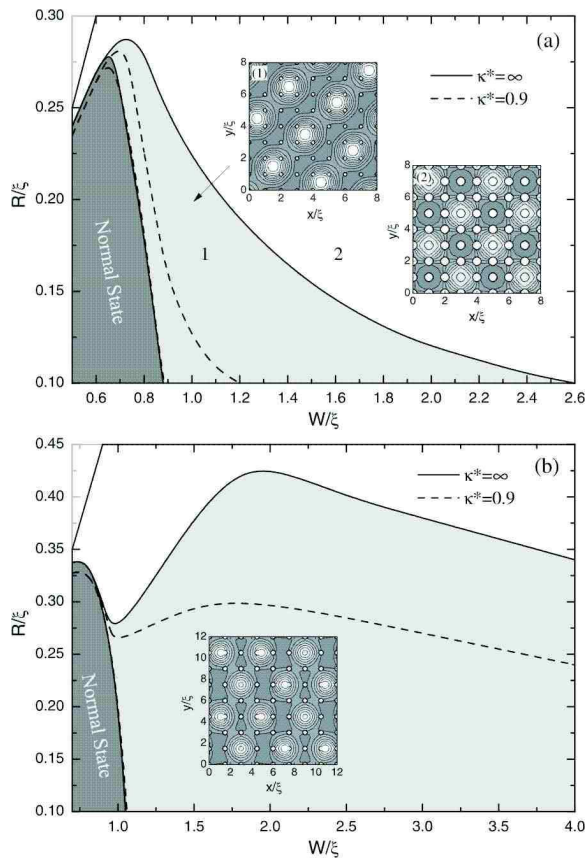}
\vspace{0cm} \caption{The phase diagram: square (white region) and
deformed triangular (light gray region) vortex lattice as a
function of the radius $R$ and period $W$ of antidots at
$H=H_{1/8}$(a) and $H=H_{3/16}$(b)  for $\kappa^*=\infty$ (solid
curves) and $\kappa^*=0.9$ (dashed curves). The insets show the
Cooper-pair density plots of the corresponding states.}
\label{fig.11}
\end{figure}

It is well known that the regular triangular vortex lattice has
the lowest energy in superconductors with no pinning.\cite{abri_2}
As we have shown above the square lattice of pinning sites impose
its own symmetry on the vortex structure. If the vortex-pinning
strength in a periodic square array is reduced, the vortex-vortex
repulsion starts to dominate over the pinning force and the
triangular lattice is recovered. Transition between these phases
was recently studied in Ref.~\cite{pogosov} as a function of the
amplitude of the vortex-pinning site interaction and the
characteristic length scale of this interaction within the London
theory (\textit{i.e.} $\kappa^*=\infty$). They showed that the
transition between triangular and square vortex lattice occurs for
increasing strength of the pinning potential in the case of small
values of pinning potential length scale. In Ref.~\cite{pogosov} a
model periodic pinning potential was introduced, the parameters of
which are difficult to relate to any growth parameters of the
sample.

\begin{figure} \centering
\vspace{0cm}
\includegraphics[scale=0.5]{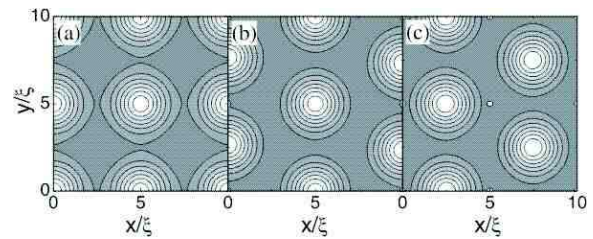}
\vspace{0cm} \caption{Contour plot of the Cooper-pair density for
different vortex states at $H=H_1$. The radius of the antidots is
$R=0.14\xi$, the period is $W=5\xi$ and effective GL parameter is
$\kappa^*=\infty$.} \label{fig.12}
\end{figure}

To circumvent the latter problem we studied the case of weak
pinning potential by introducing small antidots. Thus in order to
decrease the pinning force in our calculations we just reduced the
radius of the antidots $R$ for a given period of antidot lattice
$W$. Calculations are done for a $8\times 8$ unit cell
($W_x=W_y=8W$) with grid points $256\times256$. Fig.~\ref{fig.11}
shows the phase diagram: the transition between the pinned (white
region) and triangular (light gray region) vortex lattice as a
function of the radius $R$ and period $W$ of antidots for the
applied fields $H=H_{1/8}$(a) and  $H=H_{3/16}$(b). Let us first
discuss the results for $H=H_{1/8}$. When the pinning strength is
small (small $R$), it is energetically favorable to form a
triangular lattice, where vortices are located between the holes
[left inset in Fig.~\ref{fig.11}(a)]. For larger radius of the
holes a square vortex lattice becomes the ground state [right
inset in Fig.~\ref{fig.11}(a)]. The critical radius of the holes
$R$ to pin the vortices decreases with increasing period, contrary
to the one corresponding to the case of two-flux quanta captured
by the holes [see Fig.~\ref{fig.8}(b)]. If we decrease the GL
parameter $\kappa$ the transition between pinned and triangular
vortex lattices decreases (dashed curves in Fig.~\ref{fig.11}) due
to the short range interaction between the vortices.

In Ref.~\cite{pogosov} the phase diagram for the transition
between the triangular and square vortex lattice was found to be
the same for all submatching fields not exceeding $H=H_1$, except
$H=H_1$ and $H=H_{1/2}$. Contrary to this results our calculations
give a different phase diagram for different fractional matching
fields. As an example, we show in Fig.~\ref{fig.11}(b) the
transition lines between triangular and square pinned vortex
configurations for $H=H_{3/16}$. The triangular vortex lattice is
formed where some of the vortices are pinned and some of them are
located between the antidots [see the inset of
Fig.~\ref{fig.11}(b)]. For this value of the field the triangular
vortex lattice is found for larger values of the period $W$ and
radius $R$ of the antidots.

For the fields $H=H_{1}$ and $H=H_{1/2}$ the pinned vortex lattice
has a square symmetry. An intermediate vortex configuration for
these fields was obtained in Ref.~\cite{pogosov} where vortices in
odd rows of pinning centers are depinned and are located between
the pinning sites forming a kind of triangular lattice. Our
calculations show that such vortex configurations can be found
only as a metastable state and for relatively larger values of the
period ($W>4\xi$).  To make this more clear we plot in
Fig.~\ref{fig.12} the ground state (a) and metastable (b,c) vortex
configurations at the first matching field. In addition to the
triangular vortex state given in Ref.~\cite{pogosov} [see
Fig.~\ref{fig.12}(b)] another triangular vortex state is found,
where all the vortices are located in the interstitial region
[Fig.~\ref{fig.12}(c)].



\section{Critical current of patterned SC films}

In the previous sections we showed that vortex configurations that
are commensurate with the periodic arrays of antidots exhibits
well-defined matching phenomena, which leads to pronounced peaks
in the critical current (see for example Ref.~\cite{MoshPRB98}).
However, the stability of these vortex states strongly depend on
the parameters of the sample. For example, a multi-quanta vortex
state become energetically favorable for large radius of the
holes, while small holes can capture only a single vortex. The
additional vortices located in interstitial sites reduces the
critical current considerably. Therefore, we first investigate the
critical current of our sample as a function of the relevant
antidot parameters.

The first step to calculate the critical current is to accurately
determine the vortex ground state for given applied magnetic
field, in a manner described in previous section. Then the applied
current in the $x$ direction is simulated by adding a constant
$A_{cx}$ to the existing vector potential of the applied external
field.~\cite{mishko} With increasing $A_{cx}$ we find a critical
value of $A_{cx}$ such that a stationary solution to Eqs.
(\ref{GL1}-\ref{GL2}) cannot be found since a number of vortices
is driven in motion by the Lorentz force. The current $j_{x}$ in
the sample corresponding to the given value of $A_{cx}$ is
obtained after integration of the $x$-component of the induced
supercurrents in the $y$-cross-section. The maximal achievable
value of $j_{x}$ denotes the critical current $j_{c}$.

\begin{figure} \centering
\vspace{0cm}
\includegraphics[scale=0.45]{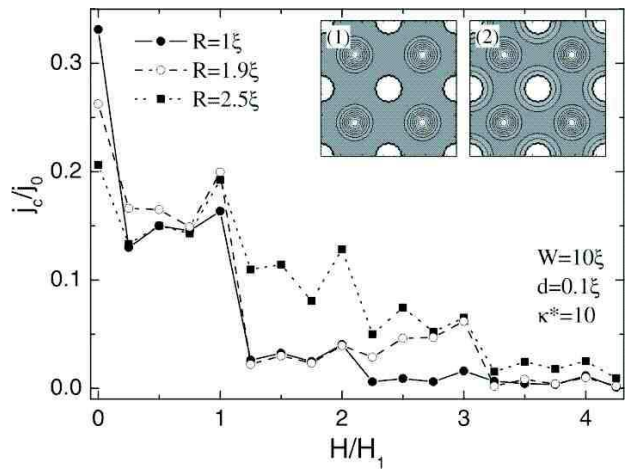}
\vspace{0cm} \caption{Critical current density (in units of
$j_{0}=cH_{c2}\xi/4\pi\lambda^{2}$) as a function of the applied
magnetic field (in units of the first matching field $H_{1}$) for
three values of the antidot radius: $R=1\xi$ (solid circles),
$R=1.9\xi$ (open circles) and $R=2.5\xi$ (squares). The antidot
lattice period is $W=10\xi$, the film thickness is $d=0.1\xi$ and
the effective GL parameter is $\kappa^{*}=10$. The insets show the
contour plots of the Cooper-pair density at the second (1) and
third (2) matching fields for $R=1.9\xi$.} \label{fig.13}
\end{figure}
\begin{figure} [t] \centering
\vspace{0cm}
\includegraphics[scale=0.45]{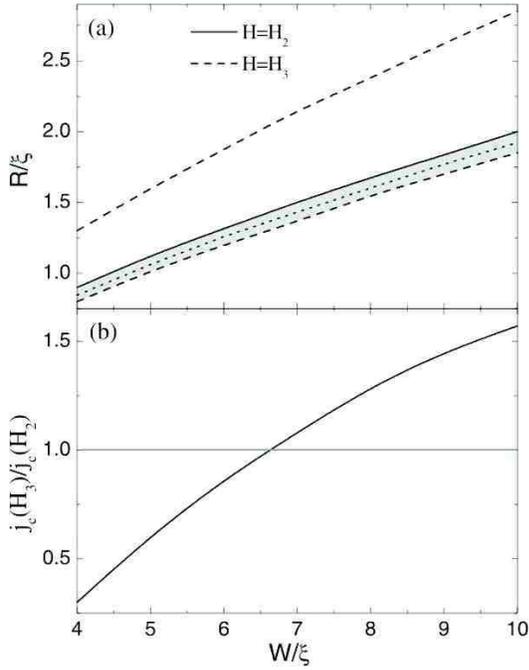}
\vspace{0cm} \caption{(a) The dependence of the antidot occupation
number $n_o$ as a function of antidot radius $R$ and lattice
period $W$ (see also Fig.~\ref{fig.7}). The solid (dashed) line
indicates the transition between the states with different $n_o$
at $H=H_2$ ($H=H_3$). The shadowed area indicates the vortex state
with a single interstitial vortex for both applied fields (the
occupation number in this region is $n_{o}=1$ for $H=H_2$ and
$n_{o}=2$ for $H=H_3$). (b) The ratio of $j_{c}(H_3)/j_{c}(H_2)$
as a function of the period $W$, where the radius of the antidots
correspond to the dotted line in the middle of the dashed area.}
\label{fig.14}
\end{figure}

Fig.~\ref{fig.13} shows the critical current density $j_{c}$ (in
units of $j_{0}=cH_{c2}\xi/4\pi\lambda^{2}$) as a function of
applied magnetic field (normalized to the first matching field
$H_{1}$) for different values of the antidot radius $R$ for fixed
value of the antidot lattice period $W$. For small radius (solid
circles), where only one vortex can be pinned by the hole, the
peaks at the matching fields decrease with increasing applied
field. The opposite behavior is found when there is a caging
effect,~\cite{golibEPL} \textit{i.e.} $j_c(H_n)<j_c(H_{n+1})$,
which \textit{e.g.} is found for radius $R=1.9\xi$ (open circles)
for $n_o=1$ and $n_o=2$. This effect occurs when there are the
same number of interstitial vortices but the number of pinned ones
are different at the different matching fields. In this case the
interstitial vortices feel a stronger repulsive interaction when
there are a larger number of pinned vortices. As is shown in
Fig.~\ref{fig.13} (open circles), a higher critical current is
found for the third matching field, when a double vortex occupies
each hole and a single one is located at the interstitial, than
for the second matching, with one vortex in each hole and a single
interstitial vortex (see the insets of Fig.~\ref{fig.13}). This
effect disappears with further increasing the radius $R$ due to
the different occupation number $n_o$, \textit{i.e.} no
interstitial vortices at $H=H_2$.

In order to show the range of radius $R$ and period $W$ of
antidots, where this caging effect is active, we constructed a
$R-W$ phase diagram for $H=H_2$ and $H=H_3$, shown in
Fig.~\ref{fig.14}(a). The shadowed area indicates the vortex state
with a single interstitial vortex for both $H=H_2$ (solid line)
and $H=H_3$ (dashed curves). Fig.~\ref{fig.14}(b) shows the ratio
$j_c(H_3)/j_c(H_2)$ as a function of period $W$. The critical
radius $R$ is taken from the middle of the region (dotted curve).
It is seen from this figure that, although we have the same vortex
structure for all values of the period $4\xi\leq W \leq 10\xi$,
the enhancement of $j_c$ is found only for $W\gtrsim 6.6\xi$. For
small period the pinned vortices at $H=H_3$ suppresses
superconductivity around the holes and interstitial vortices are
easily set into motion, reducing the critical current.

\begin{figure}\centering
\vspace{0cm}
\includegraphics[scale=0.525]{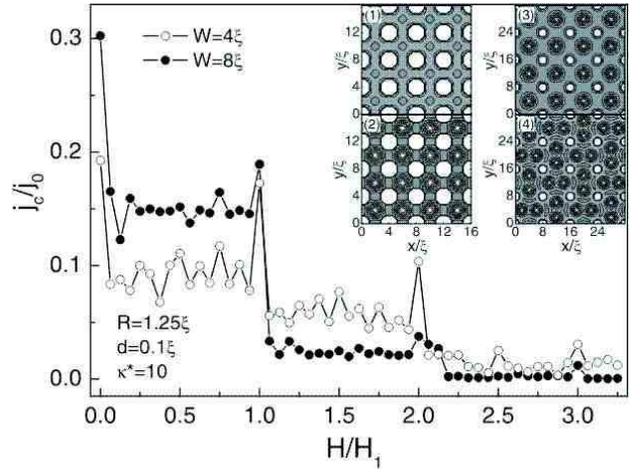}
\vspace{0cm} \caption{Critical current density of the
superconducting film as a function of the applied magnetic field
for two values of the antidot lattice period: $W=8\xi$ (solid
circles) and $W=4\xi$ (open circles). The insets show contour
plots of the Cooper-pair density at the second (1,3) and third
(2,4) matching fields for $W=4\xi$ (1,2) and $W=8\xi$ (3,4). The
radius of the holes is $R=1.25\xi$, the film thickness is
$d=0.1\xi$ and the effective GL parameter is $\kappa^{*}=10$.}
\label{fig.15}
\end{figure}

Fig. \ref{fig.15} shows the critical current density as a function
of the field for two values of the period: $W=4\xi$ (open circles)
and $W=8\xi$ (solid circles) at $R=1.25\xi$. As we showed above,
the $j_{c}(H)$ curve shows pronounced maxima at integer fields
$H_{1}$, $H_{2}$ and $H_{3}$ and at some of the fractional
matching fields. However, while the qualitative behavior of
$j_{c}(H)$ in Fig. \ref{fig.15} is as expected, its quantitative
behavior reveals a counterintuitive phenomenon. Namely, one
expects higher critical current in the sample with larger
interhole distance, simply due to the presence of more
superconducting material. Indeed, that is the case for $H\leq
H_{1}$, where the superconductor is able to compress all flux
lines in the holes. However, for higher magnetic fields, the
critical current drops sharply immediately after the first
matching field $H_{1}$, which is related to the appearance of
interstitial vortices. On the other hand, the smaller interhole
distance affect the hole occupation number (see Fig. \ref{fig.7}),
and the additional vortices after $H=H_{1}$ are still captured by
the holes (as illustrated by Cooper-pair density plots in the
inset of Fig. \ref{fig.15}). Consequently, the critical current in
this case is larger for smaller periodicity. Note that even for
smaller periodicity a sharp drop in $j_{c}$ is observed for
$H>H_{1}$, as every additional vortex disturbs the stability of
the vortex lattice. Even at $H=H_{2}$, although all vortices are
captured by the holes, the critical current is lower, due to a
stronger suppression of the order parameter around the holes
compared to the $H=H_{1}$ case. The height of the matching peaks
is decreasing with further increasing field (due to the presence
of interstitial vortices), which agrees with experiment (see
Ref.~\cite{MoshPRB98}), and these peaks strongly diminish for
higher fields as the vortex-flow overwhelms the pinning potential.

\begin{figure}\centering
\vspace{0cm}
\includegraphics[scale=0.6]{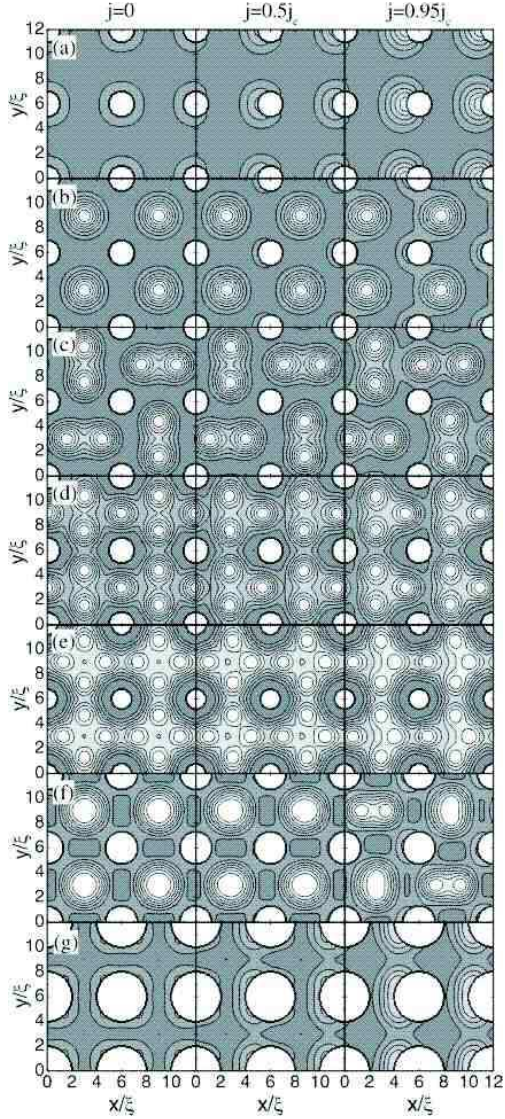}
\vspace{0cm} \caption{Contour plot of the Cooper-pair density for
$H=H_1$ (a), $H=H_2$ (b), $H=H_3$ (c,g), $H=H_4$ (d,f), and
$H=H_5$ (e), and for the applied current $j=0$ (first column),
$j=0.5j_c$ (second column) and $j=0.95j_c$ (third column). GL
parameter is $\kappa^*=10$, the period of the antidots lattice is
$W=6\xi$ and the radius of the antidots is $R=0.8\xi$ (e),
$R=1\xi$ (a-d), $R=1.3\xi$ (f) and $R=2\xi$ (g).} \label{fig.16}
\end{figure}

When we apply a \textit{dc} current into the superconductor the
vortex lattice is distorted before the vortices start moving. To
illustrate this phenomenon, we plot in Fig.~\ref{fig.16} the
Cooper pair density of the superconducting film at the applied
currents (in $y$ direction) $j=0$ (first column), $j=0.5j_c$
(second column) $j=0.95j_c$ (third column) for different matching
fields. At the first [Fig.~\ref{fig.16}(a)] and second
[Fig.~\ref{fig.16}(b)] matching fields all the vortices are
displaced over the same distance, conserving the square symmetry
in the lattice of vortices. At larger fields, when there is a
large number of interstitial vortices [Fig.~\ref{fig.16}(c-e)],
the vortex configuration is changed by the current and some of the
vortices are jammed at the interstitial sites. If we initially
have giant vortices [Fig.~\ref{fig.16}(f)] they can be split into
multivortices with increasing $j$. Our calculations also show that
there is no transition from the multivortex state to the giant
vortex state when we increase the applied current, and the
occupation number of the antidots $n_o$ is found to be independent
of $j$.

Another interesting feature following from the displacement is
found for fractional matching fields. For example,
Fig.~\ref{fig.2}(d) shows alternating two-vortex - single vortex
structure at $H=H_{5/2}$, where applying small current in
$y$-direction can shift the excess-vortex from one interstitial
site to another. Note that resulting state has identical
configuration and energy as the previous one. In order to estimate
the energy barrier between these two vortex states we performed
calculations for a superconducting film of thickness $d=13$nm with
an array of antidots with period $W=1\mu$m, radius $R=0.13\mu$m,
at temperature $T=0.9T_c$. We take $\xi(0)=40$nm and
$\lambda(0)=80$nm, which are typical values for Pb thin films. We
found an energy barrier of $\Delta F=6.2$meV, which is
significantly higher than the thermal activation energy at this
temperature ($kT=0.56$meV), but still low enough for successful
switching by a relatively weak current. Moreover, when an $ac$
current is applied to the sample, the vortex can shift back and
forth between the adjacent cells, resulting in resonant
dissipation.

%
%

\begin{figure} \centering
\vspace{0cm}
\includegraphics[scale=0.9]{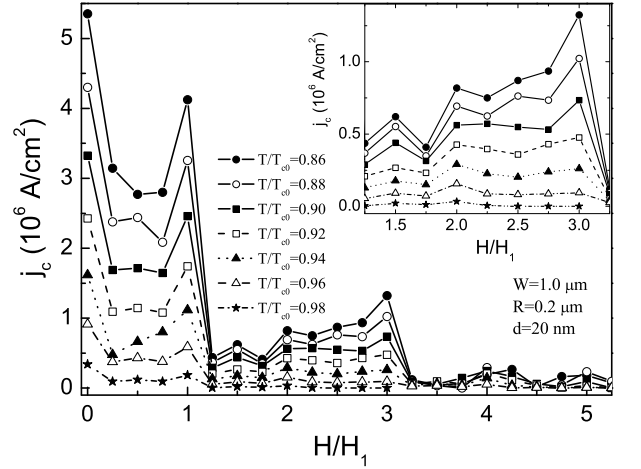}
\vspace{0cm} \caption{Critical current density of the perforated
superconducting film as a function of the applied magnetic field
(in units of the first matching field $H_{1}$) at temperatures
$T/T_{c0}=0.86\div 0.98$. The lattice period is $W=1\mu$m, the
antidot radius is $R=0.2\mu$m, and film thickness is $d=20$nm.}
\label{fig.17}
\end{figure}
\begin{figure} \centering
\vspace{0cm}
\includegraphics[scale=0.5]{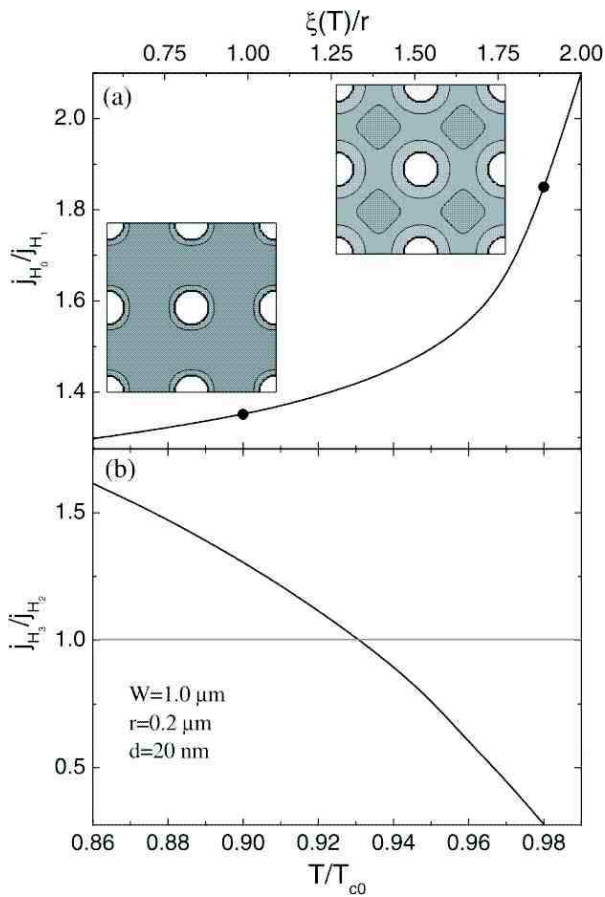}
\vspace{0cm} \caption{The ratio of $j_c(H_0)/j_c(H_1)$ (a) and
$j_c(H_3)/j_c(H_2)$ (b) as a function of temperature for the
sample with parameters given in Fig.~\ref{fig.17}. The insets show
the Cooper-pair density plots at temperatures indicated by the
solid circles in the main figure (a).} \label{fig.18}
\end{figure}

So far, we presented results at a fixed temperature. In what
follows, we include temperature in our numerical analysis through
the temperature dependence of the coherence length $\xi$. We now
consider the superconducting film with thickness $d=20$nm,
interhole distance $W=1\mu$m, and antidot radius $R=0.2\mu$m. We
choose the coherence length $\xi(0)=40$nm and the penetration
depth $\lambda(0)=42$nm, which are typical values for Pb films.
Fig.~\ref{fig.17} shows the calculated critical current of the
sample as a function of the applied field normalized to the first
matching field at temperatures $T/T_{c0}=0.86\div 0.98$. As
expected, decreasing the temperature leads to a larger critical
current for all values of the applied field. The relative height
of the peak at zero field with respect to one at the first
matching field increases with increasing temperature [see
Fig.~\ref{fig.18}(a)]). At higher temperatures, i.e. for
$\xi(T)>R$, a certain suppression of the order parameter is
present around the antidots [right inset in Fig.~\ref{fig.18}(a)]
as the core of pinned vortices overlaps with the interstitial
regions. Consequently, the suppressed order parameter leads to a
smaller $j_c$. The caging effect is found for temperatures $T\leq
0.93T_{c0}$ [see Fig.~\ref{fig.18}(b)] and it disappears with
temperature when approaching $T_{c0}$, since the vortices entirely
cover the interstitial regions and effectively destroy
superconductivity.

%

\begin{figure}[t] \centering
\vspace{0cm}
\includegraphics[scale=0.9]{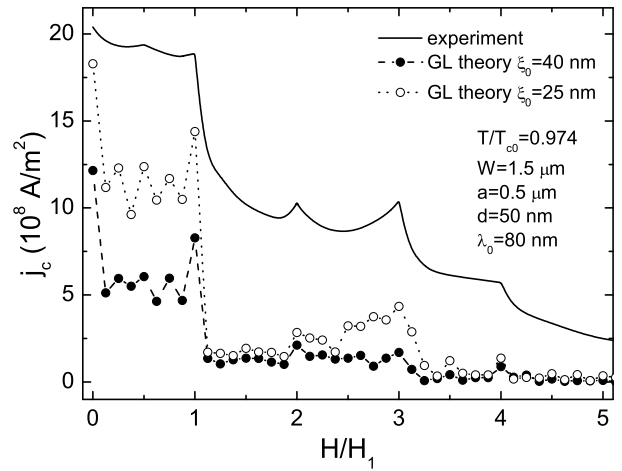}
\vspace{0cm} \caption{Numerically obtained $j_{c}(H)$
characteristics of the superconducting film with antidot arrays
(parameters given in the figure) for the coherence length at zero
temperature $\xi_{0}=40$nm (solid circles) and $\xi_{0}=25$nm
(open circles). The solid curve denotes experimental data [taken
from Ref. \cite{silha05}].} \label{fig.19}
\end{figure}

This effect that the critical current is larger for larger fields
was recently observed experimentally.~\cite{silha05} The
considered sample was a Pb film of thickness 50nm, with square
antidots of size $a=0.5\mu$m and period $W=1.5\mu$m. The coherence
length and the penetration depth at zero temperature were
estimated to be $\xi_{0}=40$nm and $\lambda_{0}=80$nm. Although
plotted for other purposes, Fig. 6(b) in Ref. \cite{silha05}
demonstrates a clear overshoot of the critical current at $H=H_3$
with respect to the one at $H=H_2$, at the temperature
$0.974T_{c0}$. Fig. \ref{fig.19} shows the comparison of the
calculated critical current density (dots) with experiment (solid
line). Our $j_c(H)$ curve shows the same qualitative behavior as
the experimental one, though a quantitative agreement is lacking
for the experimentally estimated values of $\xi_{0}$ and
$\lambda_{0}$. Better correspondence was achieved for smaller
values of $\xi_{0}$, indicating somewhat ``dirty'' sample in the
experiment. No further attempts were made to improve the
quantitative agreement with experiment because of the different
determination of $j_{c}$ in the experiment and in our theory. In
our calculations we use a dynamical criterium, \textit{i.e.} we
assume normal state as soon as vortices are set in motion, whereas
in transport measurements a certain value of the threshold voltage
was used to determine the critical current and the surface barrier
at the edges is important. Therefore, our result should be
considered as a lower limit to the experimental critical current.
The qualitative behavior of $j_c$ at the matching fields should
not be influenced by these facts.

\section{Superconducting/normal $T_{c}(H)$ phase boundary}

\begin{figure} \centering
\vspace{0cm}
\includegraphics[scale=0.9]{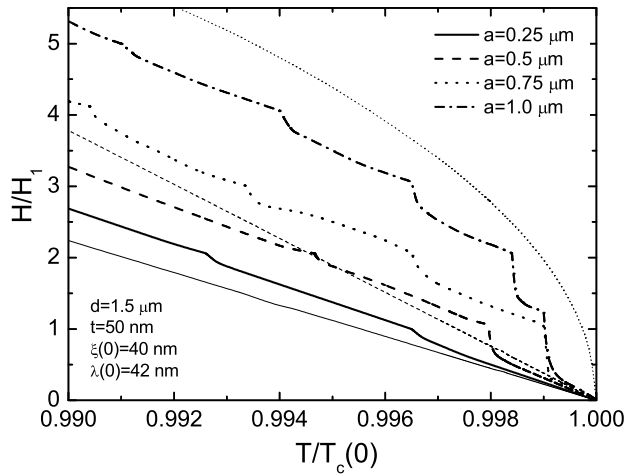}
\vspace{0cm} \caption{$H-T$ phase boundary for the superconducting
film with an antidot array. The film thickness is $d=50$nm, the
period of the antidot lattice is $W=1.5\mu$m, the antidot size is
varied as $a=0.25\mu$m (solid curve), $a=0.5\mu$m (dashed curve),
$a=0.75\mu$m (dotted curve), and $a=1.0\mu$m (dash-dotted curve).
Thin solid curve denotes the upper critical field ($H_{c2}$) of
the plain superconducting film [Eq. (\ref{Hc2})], thin dashed
curve gives the third critical field $H_{c3}=1.69H_{c2}$ for a
plain superconductor-vacuum boundary, and thin dotted curve is the
critical field of a superconducting strip with thickness
$\omega=0.5\mu$m [Eq. (\ref{Hc3})].} \label{fig.20}
\end{figure}

The presence of antidot lattice in a superconducting film not only
enhances the vortex-pinning, which was discussed in the previous
section, but also affects substantially the nucleation of
superconductivity. Due to the superconducting/vacuum interface at
the antidots, surface superconductivity will be important around
each antidot, at fields above the bulk critical field $H_{c2}(T)$.
This makes it possible to enhance the critical field in patterned
superconducting films above $H_{c2}(T)$ and even beyond the third
critical field $H_{c3}(T)$. The ratio $H_{c3}(T)/H_{c2}(T)$ tends
to the value $1.69$, the enhancement factor for a semi-infinite
slab.~\cite{gennes} However, for a dense antidot lattice a much
larger enhancement can be achieved. Namely, if the antidots are
sufficiently closely spaced, almost the entire sample may become
superconducting at high fields through surface superconductivity.

The critical field of superconducting Pb films with a square array
of antidots was investigated in Ref.~\cite{HT} by the magneto
resistance measurements. The experimentally obtained $H-T$ phase
boundary shows a cusp-like behavior with cusps at integer and some
fractional matching fields. The amplitude of the cusps depend on
the resistive criterion: the cusps become sharper and their
amplitude increases with decreasing this criterion.

We investigated numerically the $H-T$ phase boundary for a
superconducting film of thickness $d=50 \mu$m in the presence of a
regular array of square antidots with lattice period $W=1.5\mu$m.
We take the coherence length at zero temperature as $\xi(0)=40$nm
and penetration depth as $\lambda(0)=42$nm. Fig.~\ref{fig.20}
shows the calculated $T_{c}(H)$ phase diagram for different sizes
of the antidots: $a=0.25\mu$m (solid curve), $a=0.5\mu$m (dashed
curve), $a=0.75\mu$m (dotted curve), and $a=1.0\mu$m (dash-dotted
curve). For comparative reasons, we plotted also the phase
boundary for a plain film (thin solid curve) with the same
coherence length $\xi(0)=40$nm, obtained from the well-known
expression for the upper critical field
\begin{equation}
H_{c2}=\frac{\Phi_{0}}{2\pi\xi^{2}(T)}=\frac{\Phi_{0}}{2\pi\xi^{2}(0)}\Big(1-T/T_{c0}\Big).
\label{Hc2}
\end{equation}
It can be easily seen that the antidot lattice has a profound
influence on the critical magnetic field, as compared to a
reference non-patterned film. The critical temperature is enhanced
at every field, and vice versa, regardless of the size of the
antidots. Note also that matching features are present in
$T_{c}(H)$ at integer matching fields. For small radius of the
antidots matching peaks at higher integer matching fields
$H>H_{2}$ are weakly pronounced, due to the small hole-saturation
number (see Sec. III). We did not observe clear evidence of
fractional matching features.


For small radius of the antidots the sample basically acts as a
non-patterned film for temperatures close to $T_{c0}$ and the
dependence of the critical temperature on the applied field is
almost linear. For larger sizes of the antidots (e.g. dotted and
dash-dotted curves in Fig.~\ref{fig.20}), the critical field
becomes substantially higher than the third critical field of a
semi-infinite slab (thin dashed curve), and the peaks at matching
fields are more pronounced. In addition, $T_{c}(H)$ exhibits a
parabolic background as for a thin slab in a perpendicular field,
as well as to a thin film in a parallel field, which can be
described in the London limit by~\cite{tinkham}

\begin{equation}
H_{c3}=\frac{\sqrt{12}\xi(T)}{\omega}H_{c2}(T), \label{Hc3}
\end{equation}
where $\omega$ stands for the width of the superconducting strip.


%
%

\section{Conclusions}


We have studied the vortex structure of a thin superconducting
film with a regular array of antidots, which shows a rich variety
of ordered vortex lattice configurations for different matching
and fractional matching fields $H_{n}$. For small radius of the
holes, the vortex configurations with one vortex captured in each
hole and the others located in the interstitial sites are
realized, where interstitial vortices form regular patterns,
either as multi- or giant vortices, or combination of giant- and
multi-vortex states. For particular geometrical parameters of the
sample and the applied field, a symmetry imposed vortex-antivortex
configuration is found. Depending on the ratio between the hole
radius $R$ and the interhole distance $W$, multi-quanta vortices
may be forced into the antidots, in spite of their low saturation
number at smaller magnetic fields. To illustrate the transition
between possible multi-quanta states in the holes we showed a
diagram of the occupation number $n_{o}$ as a function of the
radius of the holes and interhole distance for different values of
the effective GL parameter. $n_o$ increases with decreasing
$\kappa^*$ due to the enhanced expulsion of the magnetic field
from the superconductor and giant vortices become energetically
favorable because of the attractive interaction between the
vortices.

When the pinning force of the antidots is small, \textit{i.e.}
small radius of the antidots, the triangular vortex lattice
becomes energetically favorable. Depending on the applied field
all the vortices can be located between the antidots, or some of
them are pinned by the antidots and some of them are located
between the pinning centers. We calculated the phase diagram which
shows the transition between the triangular and pinned square
vortex lattices for two values of the applied field and GL
parameter $\kappa$. We found that the results from the simple
London theory for the phase diagram are different from our GL
results for different applied fields. Moreover, we could not find
triangular vortex structures for the fields $H=H_1$ and
$H=H_{1/2}$ as a ground state.

The critical current $j_{c}$ of the sample shows well defined
peaks at different matching $H_{n}$ and fractional matching
fields, indicating that vortices are strongly pinned by antidots.
However, the level of $j_{c}$ enhancement at particular magnetic
field strongly depends on the antidot occupation number $n_o$. For
certain parameters of the sample, the critical current becomes
larger at higher matching fields, contrary to conventional
behavior.

We also studied the $T_{c}(H)$ phase boundary of regularly
perforated superconducting film. When an antidot array is present
the critical temperature $T_{c}(H)$ is enhanced compared to a
non-patterned film and distinct cusps in the phase boundary are
found for different matching fields, which is in agreement with
the experimental.~\cite{HT} This behavior is in contrast to the
Little-Parks~\cite{little} like structures found in finite size
superconductors.~\cite{golib} The increase of the antidot size for
given lattice period leads to the change of the $T_{c}(H)$
background from linear to parabolic behavior except for $T$ near
$T_{c0}$.

\section*{ACKNOWLEDGEMENTS}

This work was supported by the Flemish Science Foundation
(FWO-Vl), the Belgian Science Policy (IUAP), the JSPS-ESF NES
program, and the ESF-AQDJJ network.

\end{document}